\documentclass[aps,prc,groupedaddress,twocolumn,showpacs,amsmath,amssymb]{revtex4}
\usepackage{graphics}
\usepackage{dcolumn}
\usepackage{longtable}

\begin{document}

\title{Isomeric cross sections of fast-neutron induced reactions on $^{197}$Au}

\author{M.~Avrigeanu}
\author{V.~Avrigeanu}
\email{Vlad.Avrigeanu@nipne.ro}
\affiliation{Horia Hulubei National Institute for Physics and Nuclear Engineering, P.O. Box MG-6, 077125 Bucharest-Magurele, Romania}

\author{M. Diakaki}
\author{R. Vlastou}
\affiliation{Department of Physics, National Technical University of Athens, Greece}

\begin{abstract}
Recent accurate data obtained for the isomeric cross section of the $^{197}$Au$(n,2n)$ reaction provide a valuable opportunity to consider the question of the effective moment of inertia of the nucleus within a local consistent model analysis of all available reaction data for the $^{197}$Au target nucleus. Thus, a definite proof of a moment of inertia equal to that of the rigid--body has been obtained for $^{196}$Au nucleus while an inference of the half rigid--body value is suggested for the $^{194}$Ir nucleus. The usefulness of further measurements at incident energies up to $\ge$40 MeV has also been proved.
\end{abstract}


\maketitle

{\it I. Motivation.} 
The cross sections for nuclear reactions induced by neutrons below 20 MeV are generally considered to be reasonably well known in spite of many reactions for which the data are either conflicting or incomplete even around 14 MeV. This is the reason why new sets of accurate measured cross sections still below 20 MeV are highly desirable. One may thus understand the model constraints that are responsible for the calculated cross section variations at (i) incident energies below 20 MeV, where the statistical model (SM) calculations are most sensitive to the parameters related to residual nuclei and emitted particles, as well as (ii) above 20-30 MeV, where the pre-equilibrium emission (PE) becomes the prevailing process so that the measured data analysis may better validate the corresponding model assumptions. 

Among the former category of SM parameters some of the most important concern the nuclear level density and its spin distribution determined by the effective moment of inertia of the nucleus. This matter has been investigated for nuclei in the transitional region from well deformed to spherical nuclei near the Z=82 shell closure (e.g., Refs. \cite{np07,at11,msu11}). Former trials (e.g., \cite{vs04}) focused on the analysis of the isomeric cross section ratio in order to check the adoption of a variable moment of inertia between the half and 75\% of the rigid-body value $I_r$, for the excitation energies from the nucleus ground state (g.s.) to the nucleon binding energy, and next to $I_r$ around the excitation energy of 15 MeV \cite{va02}. The results showed that even the largest related change is still of the same magnitude with the uncertainties associated with the decay schemes and spread of the experimental data, so that a definitive conclusion on this point was precluded. However the new quite accurate data obtained for the isomeric cross section of the $^{197}$Au$(n,2n)$ reaction \cite{at11} provides a new opportunity which makes the object of the present work. Preliminary results and details of the model parameters are given elsewhere \cite{va11}. 

{\it II. Nuclear models and parameters.} 
In order to avoid the usual question marks associated with the model calculations which combine PE with equilibrium decay of the remaining compound nucleus, we have analyzed the activation cross sections of the $^{197}$Au target nucleus using a consistent local parameter set, established on the basis of various independent data. An updated version of the STAPRE-H95 code \cite{mu81,ma95} has also been used, including a generalized Geometry-Dependent Hybrid (GDH) model \cite{mb73} for PE processes, that takes into account the angular--momentum conservation \cite{ma88} and $\alpha$--particle emission while a pre--formation probability $\varphi$ \cite{eg81} is assumed with a 0.25 value. The same optical model potential (OMP) and nuclear level density parameters were  used beyond the OM and SM framework for the calculation of the PE model intra--nuclear transition rates and single--particle level (s.p.l.) densities at the Fermi level \cite{mb73,ma94,ma98}, respectively. 

\begingroup
\squeezetable
\begin{table*} 
\caption{\label{densp} Low-lying level number $N_d$ up to excitation energy $E_d$ \protect\cite{ensdf} used in cross-section calculations, and the levels and $s$-wave neutron-resonance spacings $D_0^{\it exp}$ in the energy range $\Delta$$E$ above the separation energy $S$, for the target-nucleus g.s. spin $I_0$, fitted to obtain the BSFG level-density parameter {\it a} and g.s. shift $\Delta$, for a spin cutoff factor calculated with the rigid-body value for the nucleus moment of inertia, and reduced radius $r_0$ = 1.25 fm.}
\begin{ruledtabular}
\begin{tabular}{cccccccccr} 
Nucleus   &$N_d$&$E_d$& \multicolumn{5}{c}
                  {Fitted level and resonance data}& $a$ & $\Delta$\hspace*{3mm}\\
\cline{4-8}
           &  &      &$N_d$&$E_d$&$S+\frac{\Delta E}{2}$&
                                     $I_0$&$D_0^{\it exp}$ \\ 
           &  & (MeV)&   & (MeV)& (MeV)&  &(keV)&(MeV$^{-1}$) & (MeV) \\ 
\hline
$^{190}$Ir&26&0.287&59&0.486&      &   &           & 20.40&-1.19 \\
$^{191}$Ir&20&0.686&20&0.686&      &   &           & 19.60&-0.63 \\
$^{192}$Ir&28&0.235&35&0.284& 6.198&3/2& 0.0025(3) & 20.40&-1.31 \\
$^{193}$Ir&29&0.874&29&0.874&      &   &           & 19.46&-0.62 \\
$^{194}$Ir&36&0.489&36&0.489& 6.067&3/2& 0.0058(5) & 20.00&-1.04 \\
$^{193}$Pt&32&0.701&32&0.701& 6.256& 0 & 0.0314(13)& 19.58&-0.79 \\
$^{194}$Pt&27&1.816&42&2.004&      &   &           & 19.00& 0.41 \\
$^{195}$Pt&28&0.695&28&0.695& 6.106& 0 & 0.082(10) & 18.30&-0.86 \\
$^{196}$Pt&45&2.013&54&2.093& 7.922&1/2& 0.018(3)  & 18.32& 0.33 \\
$^{197}$Pt&23&0.797&27&0.859& 5.847& 0 & 0.35(10)  & 16.65&-0.85 \\
$^{193}$Au&22&1.284&22&1.284&      &   &           & 19.50&-0.05 \\
$^{194}$Au&12&0.619& 4&0.245&      &   &           & 20.00&-0.97 \\
$^{195}$Au&36&1.443&36&1.443&      &   &           & 18.80&-0.12 \\
$^{196}$Au&56&0.573&56&0.573&      &   &           & 19.00&-1.17 \\
$^{197}$Au&14&0.948&14&0.948&      &   &           & 17.00&-0.45 \\
$^{198}$Au&30&0.573&30&0.573& 6.515&3/2& 0.0155(8) & 17.50&-1.13 \\
\end{tabular}	 
\end{ruledtabular}
\end{table*}
\endgroup

The comparison of various calculations, including  their sensitivity to model approaches and parameters, has concerned all activation channels for which there are measured data. Thus the use of model parameters that may be improperly adjusted to take into account properties peculiar to specific nuclei in the decay cascade was avoided.

The nucleon optical potential of Koning and Delaroche \cite{KD03} was found to  describe adequately \cite{va11} the RIPL-3 recommendations for the low--energy neutron scattering properties \cite{RIPL3} as well as the  recent neutron total cross sections \cite{kw06}. Actually we used the neutron transmission coefficients obtained within the code TALYS \cite{TALYS} by using RIPL 1464 potential segment. The same TALYS calculation has been used to obtain the fraction of the neutron reaction cross section corresponding to the collective inelastic scattering cross sections. Typical ratios of the direct inelastic scattering to the total reaction cross sections in the incident energy range from 4 to 40 MeV decrease from $\sim$20 to 3\%, being used for the corresponding decrease of the latter within the rest of reaction cross section calculations.

For calculation of the $\alpha$-particle transmission coefficients we have used the optical potential established previously \cite{va94} for emitted $\alpha$-particles, and supported recently by semi--microscopic analysis for $A$$\sim$90 nuclei \cite{ma06}. 

The modified energy--dependent Breit--Wigner (EDBW) model \cite{dgg79,ma87} was used for the electric dipole $\gamma$-ray strength functions $f_{E1}(E_{\gamma})$ requested for calculation of the $\gamma$-ray  transmission coefficients. The systematic EDBW-model correction factor F$_{SR}$ has been chosen in order to provide $f_{E1}(E_{\gamma})$ values close to the related experimental data and former calculations (Refs. \cite{sj86,mu91}). At the same time we used the $\gamma$-ray strength function $f_{M1}$ parameters of Ref. \cite{mu91} as well as global estimations \cite{chj77} of the $\gamma$-ray strength functions for the other multipoles $\lambda$$\le$3. The corresponding strength functions have finally been checked within the calculations of capture cross sections of $^{197}$Au nucleus in the neutron energy range from keV to $\sim$8 MeV \cite{va11}, using also the OMP and nuclear level density parameters described below. 

The nuclear level densities were derived on the basis of the back-shifted Fermi gas (BSFG) formula \cite{hv88}, for the excitation energies below the neutron-binding energy, with the parameters $a$ and $\Delta$ \cite{va02} obtained by fit of the recent experimental low-lying discrete levels \cite{ensdf} and $s$--wave nucleon resonance spacings $D_0$ \cite{RIPL3}. Above the neutron binding we took into account the washing out of shell effects \cite{avi75,arj98} using the method of Koning and Chadwick \cite{ajk97} for fixing the appropriate shell correction energy. In order to have a smooth connection we chose a transition range from the BSFG formula description to the higher energy approach, between the neutron binding energy and the excitation  energy of 15 MeV. Concerning the level density spin distribution, we used firstly a variable ratio $I/I_r$ of the nuclear moment of inertia to its rigid-body value, between  0.5 for  ground  states, 0.75 at the neutron binding energy, and 1 around  the excitation energy of 15 MeV. The results below proved however the need to consider a constant ratio $I/I_r$, equal with either 1 or 0.5. Therefore we did the fit of low--lying discrete levels \cite{ensdf} and $s$-wave nucleon resonance spacings $D_0$ also for these constant $I/I_r$ values in the range 187$<A<$206. The three sets of the $a$-parameter values obtained in this way are shown elsewhere \cite{va11} while the values for the case $I/I_r$=1 are given in Table~\ref{densp}. For the nuclei without resonance data we applied the smooth-curve method \cite{chj77} by using average $a$ values, given by the narrow $A$--range systematics, for the fit of the low--lying discrete levels.

Concerning the particle-hole state density (PSD) that plays the same role for the PE description as the nuclear-level density for SM calculations, a composite formula \cite{ma98} was used within the GDH model, with no free parameters except for the $\alpha$-particle state density $g_{\alpha}$=$A/10.36$ MeV$^{-1}$ \cite{eg81}. The PSD most important correction for the nuclear potential finite-depth was obtained by using the Fermi energy value $F$=37 MeV \cite{ab69}, while formerly \cite{va11} the value $F$=40 MeV \cite{mb73} was adopted. A linear energy dependence was adopted for the s.p.l. density $g$ of the PE excited particles, at the same time with the Fermi Gas Model (FGM) form for the exciton-configuration hole density $g_h$ (\cite{ma98} and Refs. therein).

\begin{figure}[t]
\resizebox{1.0\columnwidth}{!}{\includegraphics{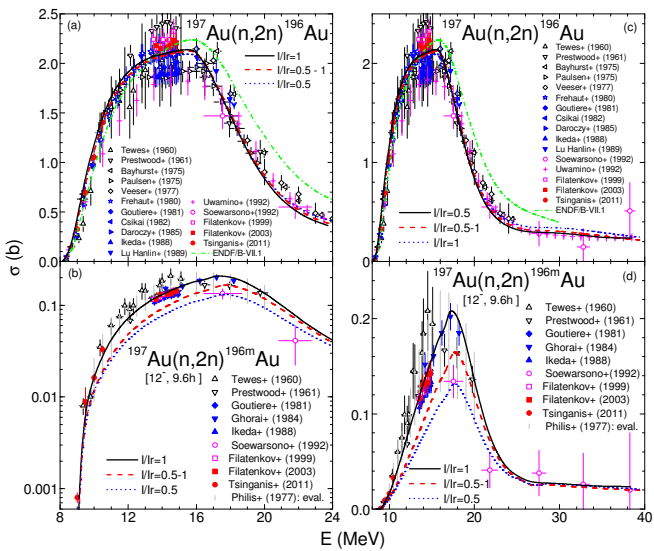}}%
\caption{\label{Au197n2n}(Color online) Comparison of experimental \cite{at11,aaf03,EXFOR}, ENDF/B-VII.1 evaluated \cite{mbc06}, and calculated cross sections of the reactions (a,c) $^{197}$Au$(n,2n)$$^{196}$Au and (b,d) $^{197}$Au$(n,2n)$$^{196m}$Au (with the corresponding spin, parity and lifetime shown between square brackets), for incident energies up to (a,b) 24 MeV and (c,d) 40 MeV, by using the excitation-energy variable ratio $I/I_r$ of the nuclear moment of inertia to its rigid-body value (dashed curve), or the constant ratios 0.5 (dotted) and 1 (solid). }
\end{figure}

{\it III. Results and discussion.}
The calculated cross sections for the $^{197}$Au$(n,2n)$$^{196}$Au reaction are in good agreement with the measured data (Fig.~\ref{Au197n2n}) especially concerning the most recent and accurate experiment \cite{at11} as well as the ones within the last decade \cite{aaf03}. However, the sensitivity of these calculations to the three options above-mentioned for the nuclear moment of inertia is so low that no conclusion is possible. On the other hand, the larger spreading of data shown in Fig.~\ref{Au197n2n}(a) around the incident energy of 14 MeV, where our calculated values match the lower limit of the most recent data \cite{aaf03}, proves the usefulness of additional accurate measurements even at these energies. The same applies to the evaluated data \cite{mbc06} from $\sim$14 to 30 MeV, while our calculations are however well describing the recent data [Fig.~\ref{Au197n2n}(c)]. Nevertheless, the large error bars of the data measured above 30 MeV put the support of model calculations under question. Hence the need for  new accurate measurements.

The comparison of the calculated and experimental cross sections for the population of the high-spin second isomeric state through the $(n,2n)$ reaction is quite a different case. This isomeric state is the 55th excited state of the $^{196}$Au residual nucleus at the top of the discrete levels taken into account in the SM calculations. Thus its population comes from the side feeding and continuum decay, so that it is fully determined by the nuclear level density and $\gamma$--ray strength functions. While the latter quantities were found to be suitably considered \cite{va11}, the model sensitivity to the nuclear moment of inertia assumption shown in Fig.~\ref{Au197n2n}(b) is so large that makes possible a certain conclusion on the real ratio $I/I_r$. The high accuracy of data recently measured  \cite{at11,aaf03} leads to a value around 1.

Unfortunately the error bars in the data set available above 20 MeV [Fig.~\ref{Au197n2n}(d)] are so large $\ge$50\% that no further assessment can be concluded on either the correct moment of inertia or the key PE model quantities that become most important at these energies (e.g., Refs. \cite{va06,ma08}).

\begin{figure}[t]
\resizebox{1.00\columnwidth}{!}{\includegraphics{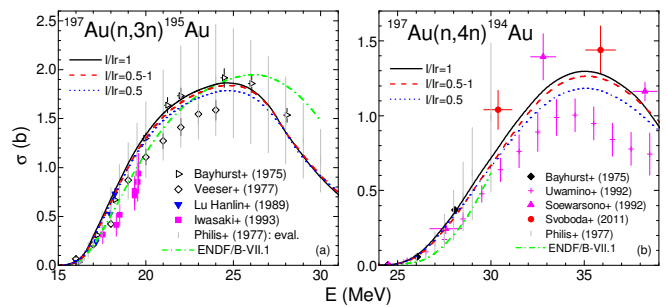}}%
\caption{\label{Au197n3n4n}(Color online) As for Fig.~\ref{Au197n2n}, but for (a) the $(n,3n)$ reaction, and (b) the $(n,4n)$ reaction on $^{197}$Au.}
\end{figure}

Similar cases of missing data with higher accuracy or better incident energy resolution are shown in Fig.~\ref{Au197n3n4n} for the $(n,3n)$ and $(n,4n)$ reactions on $^{197}$Au. On the basis of the actual measured data and model calculations, one may notice that the evaluated data accuracy has remained rather unchanged within the latest 30 years. On the other hand the need for  improved experimental data that can  establish the accuracy  of the actual phenomenological models is obvious.

\begin{figure}[t]
\resizebox{1.00\columnwidth}{!}{\includegraphics{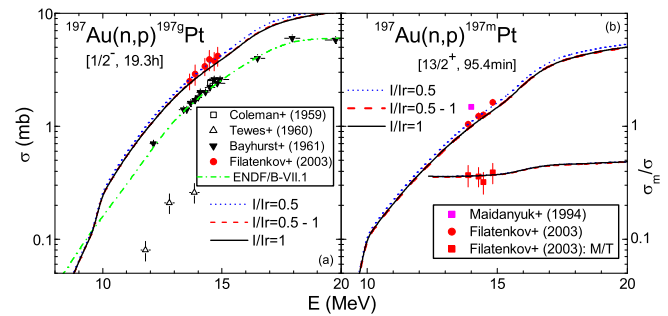}}%
\caption{\label{Au197np}(Color online) As for Fig.~\ref{Au197n2n}, but for the $(n,p)$ reaction and population of (a) the g.s. and (b) isomeric  state of the residual nucleus $^{197}$Pt, and (b) the same comparison for the isomeric ratio.}
\end{figure}

A must of an unitary reaction model analysis is the similar description of all measured data for various reaction channels, i.e. the $(n,p)$ and $(n,\alpha)$ reactions for the $^{197}$Au target nucleus. Moreover, well--known isomeric states are also populated through these reactions, and their study is a challenge for the present conclusions on the nucleus moment of inertia. As regards the $(n,p)$ reaction, our calculation results describe rather well the most recent experimental data \cite{aaf03} for the population of the ground and isomeric states of the residual nucleus $^{197}$Pt as well as the corresponding isomeric ratio (Fig.~\ref{Au197np}). However the related sensitivity of the $^{197}$Au$(n,p)$$^{197m}$Pt reaction cross sections is much lower than within previous case, due to the lower isomeric state spin and excitation energy, as well as different decay scheme. Actually this comparison is just pointing out the particular case of the high--spin second isomeric state of the $^{196}$Au nucleus.

\begin{figure}[t]
\resizebox{1.00\columnwidth}{!}{\includegraphics{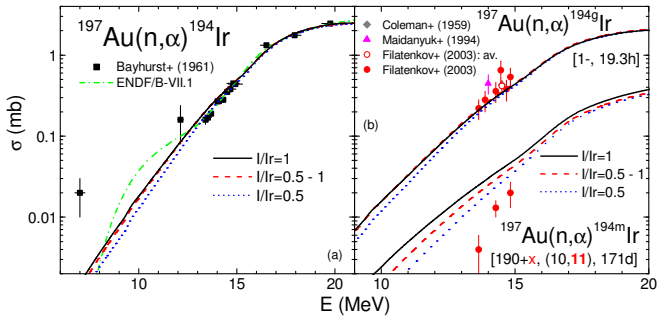}}%
\caption{\label{Au197na}(Color online) As for Fig.~\ref{Au197n2n}, but for (a) $(n,\alpha)$ reaction cross sections for the target nucleus $^{197}$Au, and (b) population of the g.s and isomeric state of the residual nucleus $^{194}$Ir.}
\end{figure}

Similar comments may apply in the case of the $(n,\alpha)$ reaction shown in Fig.~\ref{Au197na} but with an unexpected large overestimation of the isomeric cross sections measured at the same time with those for the g.s. population. Actually this isomeric state has uncertain excitation energy, between 190 and 440 keV, spin (either 10 or 11 $\hbar$) and parity \cite{bs06}. While the calculated isomeric cross sections are not sensitive to the excitation energy, assumed by us to be 440 keV, the spin value is quite important. We have considered, in agreement with the neighboring similar isotopes, the spin and parity 11$^-$. This assumption leads obviously to lower reactions cross sections, the calculated results being still larger than the measured data. The sensitivity of the calculated isomeric cross sections with respect to the option on the moment of inertia in this case is similar to the experimental errors, but the results that are closer to the measured data correspond to the ratio $I/I_r$=0.5. The considerable change of $I/I_r$ value from $^{196}$Au to $^{194}$Ir seems to be however related to results of the cranking formula \cite{kn11} and will be considered later. Since there is also a question of different OMP parameters in the incident and outgoing channels, with consequences for the present case discussed elsewhere \cite{va11}, we shall also look firstly for the answer on this difference, taking the advantage of the recent global consistent description of  $\alpha$-particle OMP in the incident channel \cite{ma10}. A comparison with the exact corresponding value $I/I_r$=0.31$\pm$2 \cite{msu11} is not yet possible since the isomeric state spin and the $\alpha$-particle OMP used in Ref. \cite{msu11} are not known.

\begin{figure}[t]
\resizebox{1.00\columnwidth}{!}{\includegraphics{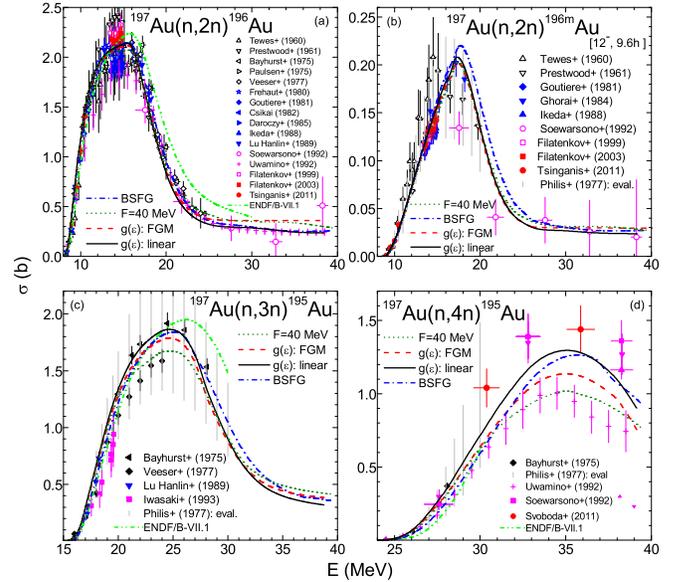}}%
\caption{\label{Au197nxn}(Color online) Comparison of experimental \cite{EXFOR}, ENDF/B-VII.1 evaluated \cite{mbc06}, and calculated $(n,xn)$ reaction cross sections for the target nucleus $^{197}$Au, by using the model assumptions and parameters given in the text (solid curves), and the alternative replacement of either the Fermi energy value $F$=40 MeV (dotted), F.G.M. energy dependence of the s.p.l. density $g$ of PE excited particles (dashed), or use of only BSFG formula for the nuclear level density (dash-dotted).}
\end{figure}

Finally, while the present local analysis has been able to provide a suitable description of most of the available data, Fig.~\ref{Au197nxn} shows that PE model assumptions could be better proved by analysis of the $(n,xn)$ reaction data above 20-30 MeV.  Nevertheless, it is obvious that improved experimental data are needed in order to establish, e.g., the correctness of the $(n,3n)$ and $(n,4n)$ reaction excitation function changes owing to different values of the Fermi energy. These changes follow the start of PE contributions due to a higher angular momentum, that happens when the corresponding local--density Fermi energies (e.g., Fig. 4 of Ref \cite{pr05}) become larger than the average excitation energy of exciton holes, within the PSD nuclear potential finite--depth correction. There are cross section variations related also to the use of either the FGM or a linear energy dependence of the s.p.l. density of excited particles within the PE exciton configurations. A description of the nuclear level density by means of BSFG formula at any excitation energy leads to eventual changes of the excitation functions at larger energies. However only few data sets, with large uncertainties, are available within this energy range. Therefore further measurements to be performed consistently at large-scale facilities as SPIRAL-2 \cite{xl06} and n\_TOF \cite{ec11}, for incident energies from threshold up to 40 as well as 100 MeV, may definitely contribute to the increase of the predictability power of actual phenomenological models. Currently no related microscopic models of a similar strength are available.

{\it IV. Conclusions.} 
Questions of consistent model analysis of all available fast-neutron reaction data for the $^{197}$Au target nucleus have been discussed within a local approach. It has thus been possible to describe most of these data, while the usefulness of further measurements to be performed at large-scale facilities, for incident energies up to 40 as well as 100 MeV, is obvious. Nevertheless, this work has shown a definite proof of a moment of inertia equal to that of the rigid--body for $^{196}$Au, by analyzing the population of its high--spin second isomeric state through the $(n,2n)$ reaction. A still open question concerns however the inference of the half rigid--body value for the $^{194}$Ir nucleus populated through the $(n,\alpha)$ reaction. Further work within the cranking formula is foreseen in order to account for the large change of the $I/I_r$ value from $^{196}$Au to $^{194}$Ir.

\bigskip
We would like to thank Mihai Mirea for useful discussions. This work was partly supported by the grant of the Romanian National Authority for Scientific Research, CNCS -– UEFISCDI, project No. PN-II-ID-PCE-2011-3-0450.

\end{document}